\documentclass[10pt,a4paper,twoside]{article}
\usepackage{epsfig}
\usepackage{baltlat6}
\usepackage{array}
\usepackage{here}
\begin{document}
\ \ \vspace{0.5mm} \setcounter{page}{1}

\titlehead{Baltic Astronomy, vol.\,25, 1--7, 2016}

\titleb{The $R_1R_2'$ outer ring revealed by young open cluster data}

\begin{authorl}
\authorb{A.~M. Mel'nik}{1*}\footnotetext{E-mail: anna@sai.msu.ru}
\authorb{P. Rautiainen}{2}
\authorb{E.~V. Glushkova}{1} and
\authorb{A.~K. Dambis}{1} \end{authorl}

\begin{addressl}
\addressb{1}{Sternberg Astronomical
Institute, Lomonosov Moscow State University, \\ Universitetskij
pr. 13, Moscow 119899, Russia;}

\addressb{2}{Department of Astronomy, University of Oulu, \\ P.O. Box 3000, FIN-90014 Oulun
yliopisto, Finland}
\end{addressl}

\submitb{Received: 2015 October 15; accepted: 2015 December 15}

\begin{summary}
The distribution of young open clusters in the Galactic plane
suggests the existence of the outer ring $R_1R_2'$ in the Galaxy.
The solar position angle $\theta_b$ providing the best  agreement
between the observed and model distribution is $\theta_b=35\pm
10^\circ$.  We compared the  $\theta_b$ values derived from
three different catalogues of open cluster and they appear to be
consistent within the errors.
\end{summary}


\resthead{The $R_1R_2'$ outer ring revealed by young open cluster
data} {A.M. Mel'nik et al.}

\sectionb{1}{INTRODUCTION}

Open clusters are born inside  giant molecular clouds, and therefore  the
concentration of young open clusters in some regions  suggests the
presence of gas there and traces the positions of spiral arms
and Galactic rings,  which are often found to be regions of
enhanced gas density and star formation.

Studies of Galactic spiral structure  are usually  based on
the classical model suggested  by Georgelin \& Georgelin (1976),
which includes  four spiral arms with the pitch angle of $\sim
12^\circ$. The spiral arms in their schema   are quite short in
the azimuthal direction winding for $\sim180^\circ$ around the
center (Figure 11 in Georgelin \& Georgelin 1976). Since then a number of authors
further developed this model (see e.g. \ Russeil 2003;
Vall{\'e}e 2008; Efremov 2011) by making the spiral structure more
symmetrical, extending spiral arms over more than
$360^\circ$ in the azimuthal direction, and attempting to
incorporate the bar. The main achievement of this
model is that it can explain the distribution of HII regions in
the Galactic disk. However, it is difficult to find  a galaxy with
the spiral structure  which includes the bar, regular four-armed
spiral pattern, and the arms making  more than a revolution around
the center (Figure 1 in Vall{\'e}e 2008).

There is extensive  evidence  for the bar in the
Galaxy  (Benjamin et al.~2005; Cabrera-Lavers et al.~2007;
Churchwell et al.~2009; Gonz\'alez-Fern\'andez et al.~2012).  The
general consensus is that  the major axis of the bar is oriented
in the direction $\theta_b=15\textrm{--}45^\circ$ in such a way
that the end of the bar closest to the Sun lies in  quadrant I.
The semi-major axis of the Galactic bar is supposed to lie in the
range $a=3.5\textrm{--}5.0$ kpc. Assuming that its end is located
close to its corotation radius, we can estimate the bar angular
speed $\Omega_b$, which appears to be constrained to the interval
$\Omega_b=40\textrm{--}65$ km s$^{-1}$ kpc$^{-1}$. This means that
the outer Lindblad resonance (OLR) of the bar is located in the
solar vicinity: $|R_{OLR}-R_0|<1.5$ kpc.

Explaining the kinematics of young objects in the Perseus
region (see its location  in Figure 1 d)  is a serious test for
different  concepts of the Galactic spiral structure. The fact
that the velocities of young stars in the Perseus region are
directed toward the Galactic center, if interpreted in terms of
the density-wave concept (Lin, Yuan  \& Shu 1969), indicates that
the trailing fragment of the Perseus arm must be located inside
the corotation circle (CR) (Burton \& Bania 1974; Mel'nik et al.
2001; Mel'nik 2003; Sitnik 2003),  and hence imposes an upper limit
for its pattern speed $\Omega_s<25$ km s$^{-1}$ kpc$^{-1}$, which
is inconsistent with the pattern speed of the bar
$\Omega_b=40\textrm{--}65$ km s$^{-1}$ kpc$^{-1}$ mentioned above.

This contradiction disappears in the model of  a two-component
outer ring $R_1R_2'$, which reproduces well the velocities of young
stars  in the Sagittarius and Perseus regions (Mel'nik \&
Rautiainen 2009; Rautiainen \& Mel'nik 2010). This model can also
explain  the position of the Carina-Sagittarius arm with respect
to the Sun and the existence of some of the so-called tangential
directions connected with the maxima in the thermal radio
continuum, HI and CO emission (Englmaier \& Gerhard 1999;
Vall{\'e}e 2008), which in this case can be associated with the
tangents to the outer and inner rings (Mel'nik \& Rautiainen
2011).

Two main classes of  outer rings and pseudorings (incomplete rings
made up of spiral arms) have been identified: rings $R_1$
(pseudorings $R'_1$) elongated perpendicular to the bar and rings
$R_2$ (pseudorings $R'_2$) elongated parallel to the bar. There is
also a combined morphological type $R_1R_2'$, which exhibits
elements of both classes (Buta 1995; Buta \& Combes 1996; Buta \&
Crocker 1991). Simulations show that outer rings are usually
located near the OLR of the bar (Schwarz 1981; Byrd et al. 1994;
Rautiainen \& Salo 1999, 2000).

We suppose that the Galaxy includes a two-component outer ring
$R_1R_2'$. The catalogue by Buta (1995) contains several dozen
galaxies with rings $R_1R_2'$. Here are some examples of galaxies
with  the $R_1R_2'$ morphology that can be viewed  as possible
prototypes of the Milky Way: ESO 245-1, NGC 1079, NGC 1211, NGC
3081, NGC 5101, NGC 5701, NGC 6782, and NGC 7098. Their images can
be found in the de Vaucouleurs Atlas of Galaxies (Buta, Corwin \&
Odewahn 2007) at http://bama.ua.edu/~rbuta/devatlas/

The study of classical Cepheids  from the catalogue by Berdnikov
et al. (2000) revealed the existence of  "the tuning-fork-like"
structure in the distribution of Cepheids: at longitudes
$l>180^\circ$   Cepheids concentrate strongly to the  Carina arm,
while at longitudes $l<180^\circ$ there are two regions of high
surface density located near the Perseus and Sagittarius regions.
This morphology suggests that outer rings $R_1$ and $R_2$ fuse
together somewhere near the Sun. We have also found some
kinematical features in the distribution of Cepheids that
suggest the location of the Sun near the descending segment of the
ring $R_2$ (Mel'nik et al. 2015ab).

In this paper we compare the distribution of young open
clusters taken from different catalogues with the model position
of the outer rings.

\sectionb{2}{RESULTS}

There are several large catalogues of open clusters. For our study
we have chosen the catalogue by Dias et al. (2002), which provides
the most reliable estimates of distances, ages and other
parameters collected from data reported by different authors. The
new version (3.4) of this catalogue contains 627 young clusters with ages less
than 100 Myr. To show that our results do not depend on the choice
of the catalogue, we also study the samples of young open clusters
from the catalogues by Loktin et al. (1994) and Kharchenko et
al. (2013), which are considered to be homogeneous. Kharchenko et
al. (2013) determined the parameters for more than 3000 clusters
based on the stellar data from  the PPMXL (R\"oser
et al. 2010) and 2MASS (Skrutskie at al. 2006) catalogues using a special
data-processing pipeline. Loktin et al. (1994) constructed their
own system of parameters (ages, color excesses, distances, and
heavy-element abundances) for  330 open clusters from published
photoelectric UBV data.

We used the simulation code developed by H. Salo (Salo 1991; Salo \&
Laurikainen 2000) to construct  different types of models that
reproduce the kinematics of OB-associations in the Perseus
and Sagittarius regions. For comparison with observation we  chose
model 3 from the series of models with analytical bars (Mel'nik \&
Rautiainen 2009). This model has a nearly flat rotation curve with
$V_c=215$ km s$^{-1}$, the bar semi-axes are equal to $a=4.0$~kpc
and $b=1.3$~kpc. The positions and velocities of $5 \times 10^4$
model particles (gas+OB) are considered at time $\sim$1 Gyr from
the start of the simulation. We scaled  this model and turned it with
respect to the Sun to achieve the best agreement between the
velocities of model particles and OB-associations in five
stellar-gas complexes identified by Efremov \& Sitnik (1988).

\begin{figure} \centering
\resizebox{\hsize}{!}{\includegraphics{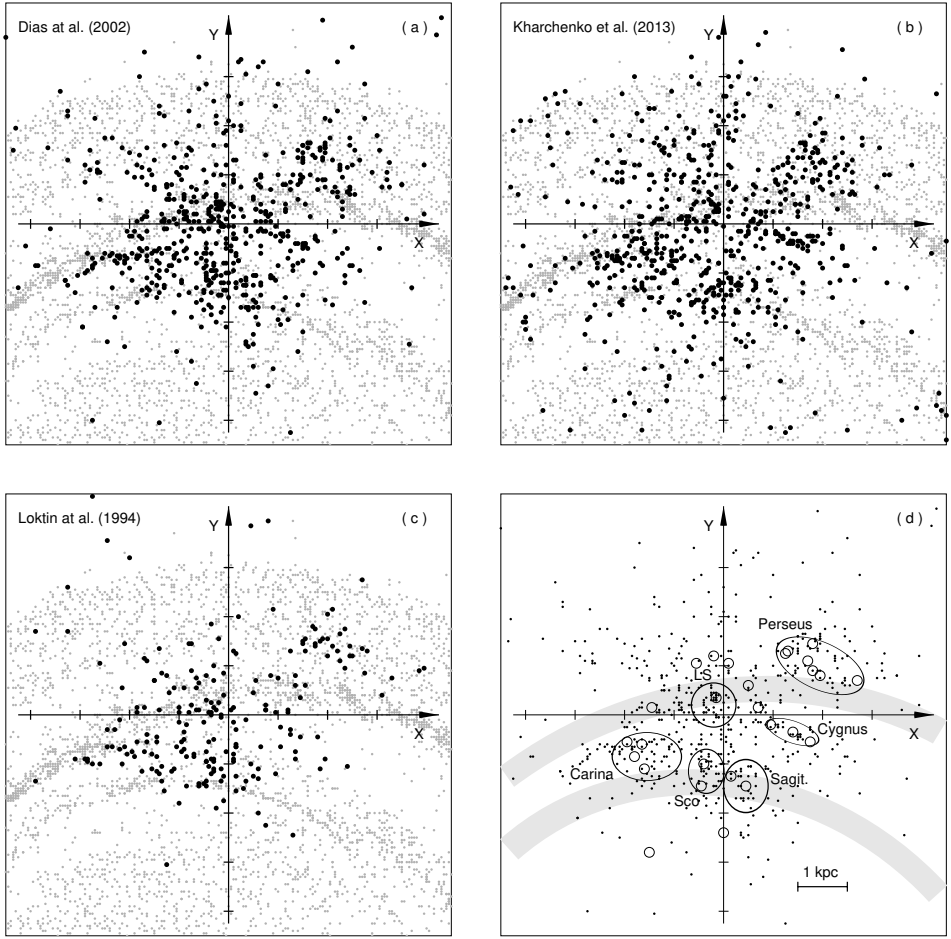}} \caption{(a)
Distribution of young  ($\log \textrm{age} <8.00$) open clusters
(black circles) from the catalogue by Dias et al. (2002) and model
particles (grey dots) in the Galactic plane. (b, c) Cluster
parameters are adopted  from the catalogues by Kharchenko et al.
(2013)  and Loktin et al. (1994), respectively. (d) The
distribution of young open clusters from the catalogue by Dias et
al. (2002) (black dots)  and rich OB-associations (white circles)
in  the Galactic plane. The locations of the outer rings $R_1$ and
$R_2$ are indicated by the gray arches. The positions of the
Sagittarius, Scorpio, Carina, Cygnus, Local System (LS),  and
Perseus stellar-gas complexes  are drawn by ellipses. The
Sagittarius, Scorpio, and Carina complexes can be associated with
the ring $R_1$. The Perseus complex, Local System, and the Carina
region are associated with the ring $R_2$. The Carina region lies
in-between the two outer rings, where they seem to fuse together.
The Sun is at the origin. The positions of model particles are
drawn for $\theta_b=45^\circ$. The $X$-axis points in the
direction of Galactic rotation and the $Y$-axis is directed away
from the Galactic center.}
\end{figure}

Figure 1 (a) shows the distribution of young  ($\log \textrm{age}
<8.00$) open clusters  from the catalogue by Dias et al. (2002)
and that of model particles  in the Galactic plane. Figure 1 (b) and (c)
show similar plots  for the  catalogues by
Kharchenko et al. (2013) and by Loktin et al. (1994),
respectively. Generally, all three distributions demonstrate
lower surface density of clusters  in quadrant III: young objects
within $r<1.5$ kpc concentrate to the Sun, while more distant
objects distribute nearly randomly over a large area. The lack of
distant objects in quadrant III is a crucial point for our
model, because the outer rings are mostly located in three other
quadrants: I, II, and IV.

Figure 1(d) shows the distribution of young open clusters
and rich OB-associations. Only OB-associations containing more
than 30 members ($N_t>30$) in the catalogue by Blaha \& Humphreys
(1989) are shown. The figure also indicates the positions of the
Sagittarius, Scorpio, Carina, Cygnus, Local System,  and Perseus
stellar-gas complexes identified by Efremov \& Sitnik (1988). We
can see that the Sagittarius, Scorpio, and Carina complexes can be
associated with the ring $R_1$. The Perseus complex, Local System,
and the Carina region can be associated with the ring $R_2$. The Carina
region lies in-between the two outer rings, where they seem to
fuse together.

The simulated distribution can be fitted by two ellipses oriented
perpendicular to each other. The outer ring $R_1$ can be
represented by the ellipse with the semi-axes $a_1=6.3$ and
$b_1=5.8$~kpc, while the outer ring $R_2$ fits well the ellipse
with $a_2=8.5$ and $b_2=7.6$~kpc. These values correspond to the
solar Galactocentric distance $R_0=7.5$ kpc. The ring $R_1$ is
stretched perpendicular to the bar and the ring $R_2$ is aligned
with the bar, and therefore the position of the sample of open
clusters with respect to the rings is determined by the position
angle $\theta_b$ of the Sun with respect to the major axis of the
bar. We now try  to find the optimum angle $\theta_b$ providing
the best agreement between the positions of the open clusters  and
the orientation of the outer rings.

Figure 2  shows the  $\chi^2$ functions -- the sums of normalized
squared deviations  of open clusters from the outer rings --
calculated for different values of the angle $\theta_b$. For each
cluster we determined the minimum distance to each of the outer rings
and then adopted the smallest of the two values. We can see that
three $\chi^2$ functions computed for data from the catalogues by
Dias at al. (2002), by Kharchenko et al. (2013), and by Loktin et
al. (1994)  reach their minima at $\theta_b=35\pm3^\circ$,
$30\pm4^\circ$, and $43\pm3^\circ$, respectively. Table~1 lists
the parameters of the observed sample: the number $N$ of clusters,
the standard deviation $\sigma$ of a cluster from the model
position of the outer rings, and the angle $\theta_b$
corresponding to the minimum on the $\chi^2$ curve. Averaging
these three values gives the mean estimate of $\theta_b=36\pm7^\circ$.

An analysis aimed at investigating eventual bias of our method
revealed   a small systematical shift of  $\pm5^\circ$: the
overdensity in the 1-kpc region  from the Sun ($r<1$ kpc)
increases  the estimated $\theta_b$, whereas objects located in
the    direction of the anticenter at $r>1.5$ kpc decrease it. The
upper limit for the uncertainty including the systematical
$\pm5^\circ$ and random $\pm5^\circ$ errors appears to be
$\pm10^\circ$ (Mel'nik et al. 2016).

\begin{figure}[t]
\centering \resizebox{10 cm}{!}{\includegraphics{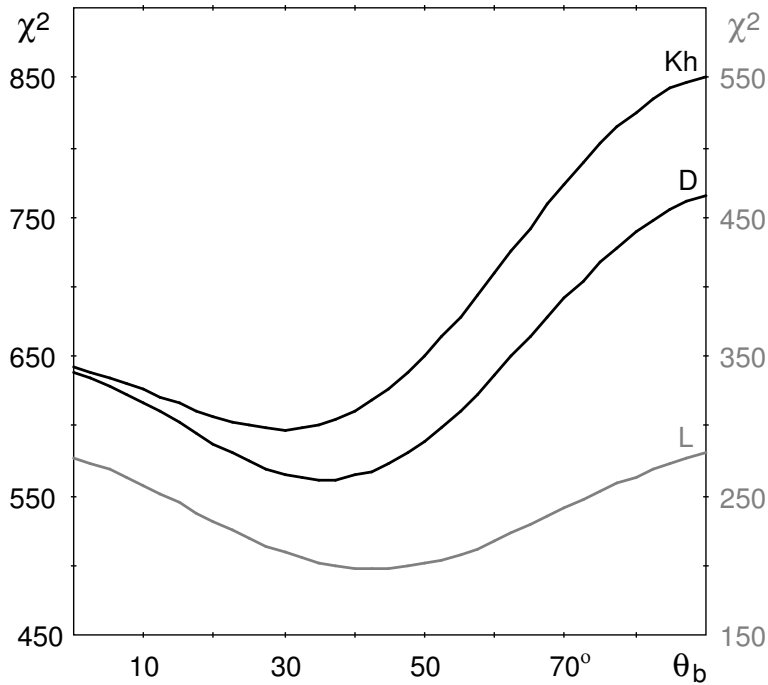}}
\caption{The $\chi^2$ functions calculated for different solar
position angles $\theta_b$ with respect  to the major axis of the
bar. The letters "D", "Kh", and "L" denote the curves calculated
for young open clusters from the catalogues by Dias et al. (2002),
Kharchenko et al. (2013),  and  Loktin et al. (1994),
respectively. The $\chi^2$ curve calculated for the data by Loktin
et al. (1994) is shown by the gray line, which is drawn for
another range of the $\chi^2$ values shown   on the right vertical
axis in gray. The three $\chi^2$ curves reach their minima at
$\theta_b=35\pm3^\circ$ (D), $30\pm4^\circ$ (Kh) and
$43\pm3^\circ$ (L).}
\end{figure}

\begin{table}
\centering \caption{Parameters of the samples}
 \begin{tabular}{lccc}
 \\[-7pt] \hline\\[-7pt]
 Catalogue & $N$   &   $\sigma$ & $\theta_b$\\
  \\[-7pt] \hline\\[-7pt]
Dias  et al. (2002)             & 564 &0.80 kpc & $35\pm3^\circ$\\
  \\[-7pt] \hline\\[-7pt]
Kharchenko  et al. (2013)       & 613 &0.81 kpc & $30\pm3^\circ$\\
  \\[-7pt] \hline\\[-7pt]
Loktin  et al.  (1994)          & 200 &0.68 kpc & $43\pm3^\circ$\\
\hline
\end{tabular}
\end{table}

\sectionb{3}{CONCLUSIONS}

We study the distribution  of young open clusters with the data adopted
from the catalogues by  Dias at al. (2002),  Kharchenko et al.
(2013), and Loktin et al. (1994) in terms of the model of the
Galactic ring $R_1R_2'$. The  solar position angle
that provides the best agreement between the distribution of observed
clusters and that of model particles appears to be $\theta_b=35^\circ$,
$30^\circ$, and $43^\circ$, respectively. It is "the
tuning-fork-like" structure in the distribution of young clusters
that determines the angle $\theta_b$ being close to 45$^\circ$. At
longitudes $l>180^\circ$ (quadrants III and IV) the two outer rings
fuse together to form one spiral fragment -- the Carina arm,
whereas at longitudes $l<180^\circ$ (quadrants I and II) the two outer
rings exist separately producing the Sagittarius and Perseus
arm-fragments.

Note that  our  study of the sample of classical Cepheids yields
$\theta_b=37\pm13^\circ$ for the position angle of the Sun with
respect to the bar's major axis (Mel'nik et al 2015ab). The
cause of this coincidence is the presence of a tuning-fork-like
structure in the Cepheid distribution as well.

We also study the distribution of  young open clusters and
OB-associations  with negative radial residual velocities $V_R$,
which within 3 kpc of the Sun must outline the descending segment
of the ring $R_2$. Clusters and OB-associations concentrate to the
fragment of the leading spiral arm (see also Mel'nik 2005),
suggesting that the Sun is located near the descending segment of
the ring $R_2$. Furthermore, the azimuthal velocity $V_T$ of these
clusters and associations decreases with increasing $x$ coordinate
-- a trend expected for objects of the ring $R_2$.  For more
details see the full paper by Mel'nik et al. (2016).

\thanks{We thank H. Salo for sharing his N-body code.
This work was supported in part by the Russian Foundation for
Basic Research (project nos.~13\mbox{-}02\mbox{-}00203,
14\mbox{-}02\mbox{-}00472), and the joint grant by the Russian
Foundation for Basic Research and Department of Science and
Technology of India (project no. RFBR 15-52-45121 -
INT/RUS/RFBR/P-219). The analysis of open cluster data was
supported by the Russian Scientific Foundation (grant no.
14-22-00041).}

\References

\refb Benjamin R. A., Churchwell E., Babler B. L. et al. 2005,
ApJ, 630, L149

\refb Berdnikov L.~N., Dambis A.~K., Vozyakova O.~V. 2000, A\&AS,
143, 211

\refb Blaha C., Humphreys R.~M. 1989, AJ, 98, 1598

\refb Burton W.~B., Bania T.~M. 1974, ApJ, 33, 425

\refb Buta  R. 1995, ApJS, 96, 39

\refb Buta R., Combes F. 1996, Fund. Cosmic Phys., 17, 95

\refb Buta R., Crocker D. A. 1991, AJ, 102, 1715

\refb Buta R., Corwin H.~G., Odewahn S.~C. 2007, The de
Vaucouleurs Atlas of Galaxies, Cambridge Univ. Press., Cambridge

\refb Byrd G., Rautiainen P., Salo H., Buta R., Crocker D. A.
1994, AJ, 108, 476.

\refb Cabrera-Lavers A., Hammersley  P. L., Gonzalez-Fernandez C.
et al. 2007, A\&A, 465, 825

\refb Churchwell E. et al. 2009, PASP, 121, 213

\refb Dias W.~S., Alessi B.~S., Moitinho A., Lepine J.~R.~D. 2002,
A\&A, 389, 871

\refb Efremov Y.N. 2011, Astron. Rep., 55, 108

\refb Efremov Y.N., Sitnik T.~G. 1988, Soviet Astron. Lett., 14,
347

\refb Englmaier P., Gerhard O. 1999, MNRAS, 304, 512

\refb Georgelin Y.~M., Georgelin Y.~P. 1976, A\&A, 49, 57

\refb Gonz\'alez-Fern\'andez C., L\'opez-Corredoira M., Am\^ores
E. B., Minniti D., Lucas P., Toledo I., 2012, A\&A, 546, 107

\refb  Kharchenko N.~V., Piskunov A.~E., R\"oser S., Schilbach E.,
Scholz R.-D., 2013, A\&A, 558, 53

\refb Lin C.~C., Yuan C., Shu F.~H. 1969, ApJ, 155, 721

\refb Loktin A.~V., Matkin N.~V., Gerasimenko T.~P. 1994, Astron.
Astrophys. Trans., 4, 153

\refb Mel'nik A.~M., Dambis A.~K.,  Rastorguev A.~S. 2001, Astron.
Lett., 27, 521

\refb Mel'nik A.~M. 2003, Astron. Lett., 29, 304

\refb Mel'nik A.~M. 2005, Astron. Lett., 31, 80

\refb Mel'nik A.~M., Rautiainen  P. 2009, Astron. Lett. 35, 609

\refb Mel'nik A.~M., Rautiainen  P. 2011, MNRAS, 418, 2508

\refb Mel'nik A.~M., Rautiainen  P., Berdnikov L.~N., Dambis
A.~K., Rastorguev A.~S. 2015a, AN, 336, 70

\refb Mel'nik A.~M., Rautiainen  P.,  Berdnikov L.~N., Dambis
A.~K.,  Rastorguev A.~S., 2015b, Baltic Astron., 24, 62

\refb Mel'nik A.~M., Rautiainen  P.,  Glushkova E.~V., Dambis
A.~K., 2016, Ap\&SS, 361, 60

\refb Rautiainen P., Mel'nik A. M. 2010, A\&A, 519, 70

\refb Rautiainen P., Salo H. 1999, A\&A, 348, 737

\refb Rautiainen P., Salo H. 2000, A\&A, 362, 465

\refb R\"oser S., Demleitner M., Schilbach E. 2010, AJ, 139, 2440

\refb Russeil D. 2003, A\&A, 397, 133

\refb Salo H. 1991,  A\&A, 243, 118

\refb Salo H., Laurikainen E. 2000, MNRAS,  319, 377

\refb Schwarz M.~P. 1981, ApJ, 247, 77

\refb Skrutskie M. F., Cutri R. M., Stiening R., et al. 2006, AJ,
131, 1163

\refb Sitnik T.~G. 2003, Astron. Lett., 29, 311

\refb Vall{\'e}e J.~P. 2008, AJ, 135, 1301

\end{document}